# Generalization of the Maier-Saupe theory to the ferroelectric nematic phase

J. Etxebarria, C.L. Folcia, and J. Ortega

*Department of Physics, Faculty of Science and Technology, UPV/EHU, Bilbao, Spain*

e-mail address: j.etxeba@ehu.es

# Generalization of the Maier-Saupe theory to the ferroelectric nematic phase

We present a generalization of the Maier-Saupe (MS) theory that incorporates the description of the ferroelectric nematic phase ($N_F$) within its scope of application. The extension of the theory is carried out in a natural way, by adding to the nematic potential of MS a term proportional to $\cos\theta$ and to the polar order parameter $<\cos\theta>$, where $\theta$ is the angle between the long molecular axis and the spontaneous polarization. The $N_F$ phase can be reached by cooling from the isotropic phase with or without the formation of an intermediate normal nematic phase, depending on the relative intensity of the polar and nonpolar terms of the nematic potential. For both phase sequences all the transitions are first order. The temperature dependence of the polar and nonpolar order parameters is calculated, and some results derived from the theory are compared with the experimental observations. It is argued that the theory is a valuable first approximation to describe some of the properties of ferroelectric nematics.

Keywords: ferroelectric nematic phase; Maier-Saupe theory; polar order parameter; mean field theory

## 1. Introduction

The Maier-Saupe (MS) theory is one of the most successful theories in the physics of nematic liquid crystals [1-3]. This theory gives a basic explanation for the existence of the nematic order and correctly describes the order of magnitude of the degree of that order. Despite its simplicity, the theory is amazingly successful in describing the nematic-isotropic (N-I) phase transition and the temperature dependence of optical anisotropy and dielectric and magnetic susceptibilities in the N phase. Obviously the theory has its own limitations. Being a mean field theory, it does not take into account the correlations in the orientations and positions between neighbouring molecules, and this is reflected in deviations of some of the theoretical predictions compared to the experimental behaviour. Furthermore, the mean field potential is described by a particularly simple model based only on a single parameter, which gives rise to further limitations for the theory. Thus, the theory predicts a universal dependence of the order parameter on the reduced temperature, which is not entirely in accordance with the observations. Anyway, the theory is a very valuable first approximation for describing many of the properties of ordinary nematics.

Recently the ferroelectric nematic phase ($N_F$) has been discovered [4-6]. That phase presents a degree of polar order in addition to the ordinary nematic order, in such a way that the head-to-tail invariance is broken and the phase exhibits spontaneous polarization along the director. Ferroelectric nematic materials present very promising potentials for various applications and can even revolutionize current nematic-based technology. Therefore, the analysis of the phase transition that gives rise to this structure seems of interest. In most materials studied so far, the $N_F$ phase is obtained from the cooling of an N phase [5,7,8] or directly from the I phase [9]. There are still not many published measurements on the behaviour of physical properties in $N_F$ phases but it is expected that this situation will change shortly.

In this paper we present a simple generalization of MS theory to describe the $N_F$ phase. This generalization is based on the introduction of a new term in the nematic potential that accounts for the observed polar order. In this regard, it is worth mentioning a recent work [10], where a free energy like the one presented here is analyzed among others, and the molecular distribution functions at the so-called critical points are examined. These points correspond to distribution functions where the free energy is a minimum or a saddle point. Interestingly, it is found that some of the distribution functions exhibit non-axisymmetric symmetries at the saddle points, though uniaxiality always holds for stable critical points. In the present work we will not consider these low-symmetry distribution functions since, being unstable, are unphysical. In fact, the main contribution of our work is to place our theoretical approach in the context of the experimental evidences observed up to now in the $N_F$ phase.

This paper is organized as follows. In the next section we will describe the model, proposing the simplest possible potential suitable for our aims. We will find the dependence with the temperature of the relevant order parameters and will give the ranges of the different phases. Then we will make a brief comparison with some experimental results. Finally we draw some conclusions.

## 2. Nematic potential

In a mean field theory it is assumed that each molecule is subjected to a field that does not depend on the orientation of the neighbouring molecules but originates from the collective alignment of all the molecules in the phase. For a normal nematic the MS model assumes a potential dependent on the polar angle $\theta$ (angle between the molecular long axis and the nematic director) in the form $V(\theta) = -V_2 \eta_2 P_2(\cos\theta)$, where $V_2 > 0$ is a constant, $P_2(\cos\theta)$ is the second Legendre polynomial, i.e., $P_2(\cos\theta) = (3\cos^2\theta - 1)/2$, and $\eta_2$ is the nematic order parameter, $\eta_2 = \langle P_2(\cos\theta) \rangle$, where the average is calculated through an angular distribution function $f(\theta)$ (independent of the azimuthal angle $\varphi$), which depends on the temperature. This independence in $\varphi$ is justified from the physical point of view, since the nematic phases are experimentally uniaxial, and also from the mathematical point of view since it can be shown [11,12] that the stable critical points are uniaxial for the nematic potential. The linear dependence of $V(\theta)$ on $\eta_2$ reflects the fact that we assume that the potential is proportional to the degree of order of the phase. On the other hand, the dependence on $\theta$ shows that the molecular orientations $\theta = 0$ and

$\theta = \pi$ are equally favoured along the **n** director, verifying then, by construction, the head-to-tail invariance.

In the case of the $N_F$ phase, such invariance is broken, the $+\mathbf{n}$ and $-\mathbf{n}$ directions being physically different. The orientation $\theta = 0$ is favoured over $\theta = \pi$, and $\eta_1 = \langle P_1(\cos\theta) \rangle = \langle \cos\theta \rangle$ is not zero. This quantity $\eta_1$, in fact, is the simplest parameter that reflects the degree of polar order in the phase. Therefore, a natural extension of the nematic potential for the case of the $N_F$ phase is the following,

$$V(\theta) = -V_2 \eta_2 P_2(\cos\theta) - V_1 \eta_1 P_1(\cos\theta), \qquad (1)$$

where $V_1 > 0$ is another constant whose value shows the importance of the polar interaction in the $N_F$ phase. For this potential, Refs. [11,12] still assure that the distribution function is independent of $\varphi$ at stable critical points.

## 3. Calculation of the free energy

In order to describe the phase transitions of a set of nematic molecules, we have to calculate the free energy per molecule in our system, $F = U-TS$, where $U$ is the internal energy and $S$ is the entropy per molecule. In terms of the distribution function these quantities are expressed as

$$U = \frac{1}{2} \int_0^\pi V(\theta) f(\theta) \sin\theta d\theta \qquad (2)$$

and

$$S = -k_B \int_0^\pi \ln[f(\theta)] f(\theta) \sin\theta d\theta, \qquad (3)$$

where the factor 1/2 in (2) is due to the fact that in the integral the interaction by pairs of molecules is added 2 times, and in (3) an "ideal gas" type entropy has been assumed.

Using (1-3) we obtain

$$F = -\frac{1}{2}V_2\eta_2^2 - \frac{1}{2}V_1\eta_1^2 + k_BT\langle \ln f \rangle . \tag{4}$$

## 4. Calculation of the angular distribution function

Equation (4) gives the $F$ function as a functional of $f(\theta)$, $F = F[f(\theta)]$. The distribution function $f(\theta)$ is calculated by requiring this function to minimize $F$ with the normalization condition

$$\int_0^\pi f(\theta)\sin\theta d\theta = 1 , \tag{5}$$

i.e.,

$$\delta\left\{F - \lambda\left[1 - \int_0^\pi f(\theta)\sin\theta d\theta\right]\right\} = 0 , \tag{6}$$

where $\lambda$ is a Lagrange multiplier.

Taking into account that

$$\delta\eta_i^2 = 2\eta_i\delta\eta_i = \int_0^\pi 2\eta_i P_i \delta f \sin\theta d\theta \quad (i = 1,2)$$

and

$$\delta\langle \ln f \rangle = \int_0^\pi \delta f (\ln f + 1)\sin\theta d\theta ,$$

we easily obtain

$$\int_0^\pi \delta f \left[-V_2\eta_2 P_2 - V_1\eta_1 P_1 + k_BT \ln f + k_BT + \lambda\right]\sin\theta d\theta = 0 ,$$

from which we deduce the expression for the distribution function

$$f(\theta) = \frac{1}{Z}\exp\left(\frac{V_2\eta_2 P_2 + V_1\eta_1 P_1}{k_BT}\right) , \tag{7}$$

being

$$Z = \int_0^\pi \exp\left(\frac{V_2\eta_2 P_2 + V_1\eta_1 P_1}{k_B T}\right)\sin\theta\, d\theta \quad . \tag{8}$$

As can be seen, the distribution function maintains the form that is deduced from MS theory, simply adding the polar contribution to the nematic potential.

## 5. Order parameters

The order parameters can be calculated by solving the self-consistent system of equations

$$\eta_1 = \int_0^\pi \frac{1}{Z}\exp\left(\frac{V_2\eta_2 P_2 + V_1\eta_1 P_1}{k_B T}\right) P_1(\cos\theta)\sin\theta\, d\theta$$

$$\eta_2 = \int_0^\pi \frac{1}{Z}\exp\left(\frac{V_2\eta_2 P_2 + V_1\eta_1 P_1}{k_B T}\right) P_2(\cos\theta)\sin\theta\, d\theta \tag{9}$$

Alternatively, $\eta_1$ and $\eta_2$ can also be found by requiring that they must minimize the free energy at each temperature. From Eq. (4), using (7) and (8) it results

$$F(\eta_1,\eta_2) = \frac{1}{2}V_2\eta_2^2 + \frac{1}{2}V_1\eta_1^2 - k_B T \ln Z(\eta_1,\eta_2) \quad . \tag{10}$$

Both the solution of the system (9) and the minimization of (10) must be done numerically. They are practically equivalent procedures for the calculation of the order parameters. We have adopted the second alternative and have verified *a posteriori* that Equations (9) are also satisfied by the solution $(\eta_1,\eta_2)$ found by minimizing $F$ in all cases.

## 6. Behaviour of $\eta_1$ and $\eta_2$ with temperature

Depending on the ratio $V_1/V_2$ we have found different behaviours for the order parameters.

i) If $V_1/V_2 <0.35$ there is an I-N-$N_F$ phase sequence. The I-N transition presents the same characteristics that are deduced from the classical MS theory:

-If $k_BT/V_2> 0.220$ the free energy is minimized for null values of $\eta_1$ and $\eta_2$ (I phase).

-If $k_BT/V_2= 0.220$ the free energy is minimized for $\eta_1= 0$ and $\eta_2= 0.43$ (N phase). The transition is first order and the behaviour of $\eta_2$ is as given by the classical theory. The latent heat per molecule is $L = V_2\Delta\eta_2^2/2$, where $\Delta\eta_2$ is the jump at the transition.

As we lower the temperature, a point is reached where free energy is minimized for $\eta_1 \neq 0$ ($N_F$ phase). Such temperature decreases on decreasing $V_1/V_2$ (see Figure 1). In all cases $\eta_1$ and $\eta_2$ show finite jumps $\Delta\eta_1$, $\Delta\eta_2 \neq 0$ when passing to the $N_F$ phase (the smaller the jumps the lower $V_1/V_2$), so that the N-$N_F$ transition always turns out to be first order, though the latent heat $L = V_1\Delta\eta_1^2/2 + V_2\Delta\eta_2^2/2$ diminishes rapidly on decreasing $V_1/V_2$. In practice $L=0$ for $V_1/V_2 <0.25$. Figure 2 shows the behaviour of the order parameters with temperature in a typical case $V_1/V_2 = 0.25$. As can be seen, $\eta_1$ grows quite quickly in comparison to $\eta_2$. $\eta_2$ also shows an anomaly at the N-$N_F$ transition, though the jump $\Delta\eta_2$ is hardly visible.

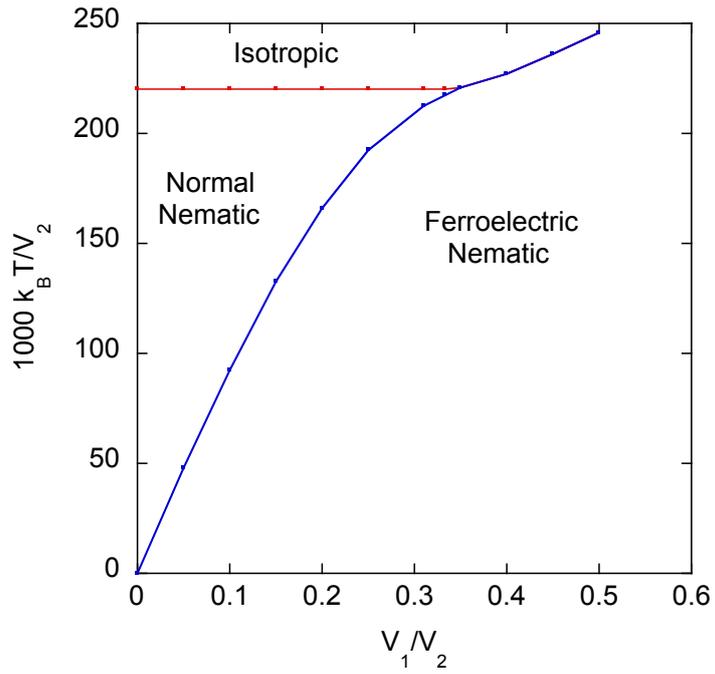

Figure 1. Phase diagram as a function of $V_1/V_2$.

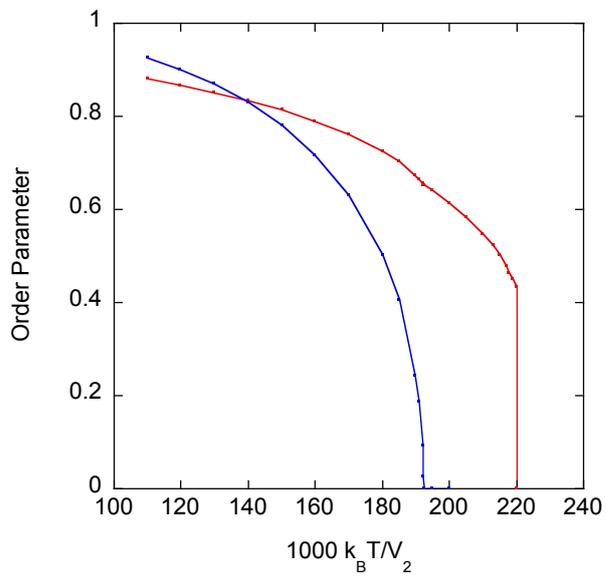

Figure 2. Temperature dependence of the order parameters $\eta_1$ (blue) and $\eta_2$ (red) for the case $V_1/V_2=0.25$.

ii) If $V_1/V_2 > 0.35$ the phase sequence is I-N$_F$, that is, both order parameters become spontaneous simultaneously. The transition temperature verifies the inequality $k_BT/V_2 > 0.220$, i.e., it is higher than the classical I-N transition temperature of MS. In all cases, the transition is strongly first order, with latent heat per molecule given by $L = V_1 \Delta\eta_1^2/2 + V_2 \Delta\eta_2^2/2$. The jumps $\Delta\eta_1$ and $\Delta\eta_2$ increase on increasing $V_1/V_2$, being $\Delta\eta_2$ always larger than the classical value 0.43. Two typical examples of this behaviour are shown in Figure 3.

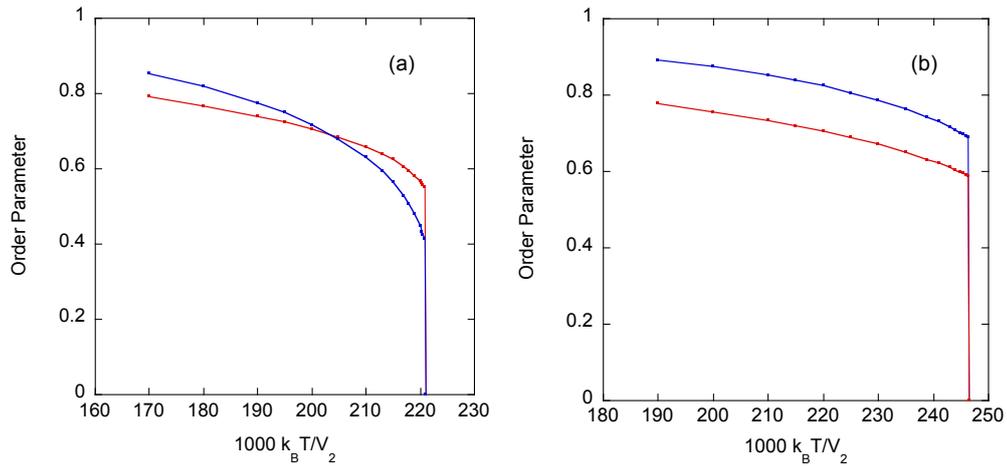

Figure 3. Temperature dependence of the order parameters $\eta_1$ (blue) and $\eta_2$ (red) for the cases $V_1/V_2=0.35$ (a) and $V_1/V_2=0.50$ (b).

### 7. Comparison with experimental results

The temperature dependence of the different physical quantities depends on the type of connection that these quantities have with the order parameters. For example, the spontaneous polarization is proportional to $\eta_1$, the birefringence and the magnetic anisotropy are proportional to $\eta_2$, while other quantities have more complicated behaviours. It is interesting to comment on some points predicted by the theory in comparison with the experimental results.

First, both the I-N-$N_F$ [5,7,8,9] and I-$N_F$ [13] sequences have been observed experimentally. In both cases, all the transitions involved have been characterized as first-order, according to the theory. The prototype material for which more measurements have been carried out (RM734) presents the first sequence, and shows a N-$N_F$ transition with a very small latent heat (0.049 kJ/mol) determined from precision adiabatic calorimetry [9]. In comparison, within our model, as $V_2=k_BT/0.22$, we predict a molar latent heat of $RT_{IN}[(V_1/V_2)\Delta\eta_1^2 + \Delta\eta_2^2]/0.44$, where $T_{IN}$ is the I-N transition temperature and $R$ the gas constant. The temperatures for the I-N and N-$N_F$ transitions are consistent with $V_1/V_2=0.25$ (see Figure 1). Taking this value and $T_{IN}=460$ K, we get a molar latent heat of the order of 0.001 kJ/mol, even less than the one observed. In contrast, the I-$N_F$ sequence presents a much larger latent heat, in agreement with our model. For example, in the compound of Ref. [13] the experimental value was 2.1 kJ/mol, whereas 2.98 kJ/mol was deduced from the theory for a typical value $V_1/V_2=0.5$ leading to the I-$N_F$ direct transition. (The necessary parameters for the calculation are plotted in Fig. 3b and $T_{I-NF}=293$ K).

Moreover, in RM734 and other similar materials:

a) The polarization follows a dependence of the type shown in Fig. 2 for $\eta_1$. The agreement is only qualitative since the measured values for $\eta_1$ show a faster growth on decreasing temperature [5,7, 14].

b) The birefringence must be proportional to $\eta_2$ and, indeed, it experimentally shows an anomaly at the N-$N_F$ transition similar to that shown in Fig. 2 for $\eta_2$ [5,7, 14].

c) The theory is capable of predicting the behaviour of other quantities with more complicated dependencies than the simple proportionality with $\eta_1$ or $\eta_2$. For example, it is known that the $N_F$ phase presents nonlinear optical activity, and it can be shown that, in a first approximation, the behaviour of the optical second-harmonic generation intensity must be proportional to $<\cos^3\theta>^2$ [15]. Obviously, this average can be easily calculated since we have the function $f(\theta)$. The result is shown in Figure 4, for the case $V_1/V_2 = 0.25$. A qualitative agreement with the experimental data is found, though, again, the experimental behaviour is steeper [8, 15].

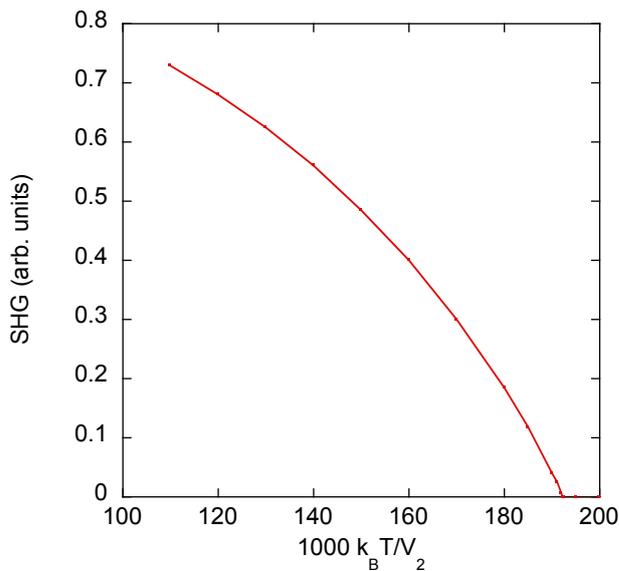

Figure 4. Temperature dependence for the optical second-harmonic generation intensity (SHG) in the $N_F$ phase for the case $V_1/V_2=0.25$.

## 8. Conclusions

We have carried out a generalization of the MS theory for the description of the $N_F$ phase by incorporating a single additional parameter in the nematic potential. In this way, the proposed generalization maintains the simplicity of the original theory. We have examined some of the predictions of the theory, finding a qualitative agreement with the experimental observations. Obviously we do not intend to make a quantitative adjustment of the results, just as the MS theory does not intend to do so either. Obviously, as in the case of the MS theory, better experimental agreement can be achieved by using more sophisticated forms of the mean field potential [16] or with improvements to the mean field approximation that take into account the influence of short-range order [17]. An example of short-range interaction leading to the $N_F$ phase has been recently given in Ref. [18], where the molecules are described as rods with surface longitudinal charge density waves. In that approach the ferroelectricity arises from the coupling of the polar order and the density of the medium. Atomistic molecular dynamics simulations [5] have also been carried out to investigate short-range molecular interactions that favour polar order, and some of these interactions have been identified. All these complicated features are integrated in the theory developed here in one single parameter, so a rigorous agreement with the observations is out of place. The theory is simply a starting point that gives a simple first approximation and, as we have seen, quite a few experimental facts are reasonably well accounted for.

Acknowledgements

This work was supported by Euskal Herriko Unibertsitatea (Project GIU18/146).

References

1. Maier W, Saupe A. Eine einfache molekulare theorie des nematischen kristallinflussigen zustandes. Z Naturforsch A (in German). 1958; 13: 564-566. DOI: 10.1515/zna-1958-0716.

2. Maier W, Saupe A. Eine einfache molekular-statistische theorie der nematischen kristallinflussigen phase 1. Z Naturforsch A (in German). 1959; 14: 882-889. DOI:10.1515/zna-1959-1005.

3. Maier W, Saupe A. Eine einfache molekular-statistische theorie der nematischen kristallinflussigen phase 2. Z Naturforsch A (in German). 1960; 15: 287-292. DOI:10.1515/zna-1960-0401.

4. Mandle RJ, Cowling SJ, Goodby JW. A nematic to nematic transformation exhibited by a rod-like liquid crystal. Phys Chem Chem Phys. 2017; 19: 11429–11435. DOI: 10.1039/C7CP00456G.

5. Chen X, Korblova E, Dong D, Wei X, Shao R, Radzihovsky L, Glaser MA, Maclennan JE, Bedrov D, Walba DM, Clark NA. First-principles experimental demonstration of ferroelectricity in a thermotropic nematic liquid crystal: spontaneous polar domains and striking electro-optics, PNAS. 2020; 117: 14021–14031. DOI: 10.1073/pnas.2002290117.

6. Mertelj A, Cmok L, Sebastián N, Mandle RJ, Parker RR, Whitwood AC, Goodby JW, Čopič M. Splay nematic phase. Phys Rev X. 2018;8: 041025. DOI: 10.1103/PhysRevX.8.041025.

7. Brown S, Cruickshank E, Storey JMD, Imrie CT, Pociecha D, Majewska M, Makal A, Gorecka E. Multiple Polar and Non-polar Nematic Phases, Chem Phys Chem. 2021. https://doi.org/10.1002/cphc.202100644.

8. Li J, Nishikawa H, Kougo J, Zhou J, Dai S, Tang W, Zhao X, Hisai Y, Huang M, Aya S. Development of ferroelectric nematic fluids with giant-ε dielectricity and nonlinear optical properties. Science Advances. 2021; 7: eabf5047. DOI: 10.1126/sciadv.abf5047.

9. Thoen J, Korblova E, Walba DM, Clark NA, Glorieux C. Precision adiabatic scanning calorimetry of a nematic-ferroelectric nematic phase transition. Liq Cryst. 2021. https://doi.org/10.1080/02678292.2021.2007550.

10. Yin J, Zhang L, Zhang P. Solution landscape of the Onsager model identifies non-axisymmetric critical points. Physica D. 2022;430:133081. DOI:https://doi.org/10.1016/j.physd.2021.133081

11. Zhou H, Wang H, Wang Q, Forest MG. Characterization of stable kinetic equilibria of rigid, dipolar rod ensembles for coupled dipole–dipole and Maier–Saupe potentials. Nonlinearity 2007;20:277-297. DOI: doi:10.1088/0951-7715/20/2/003

12. Ball JM. Axisymmetry of critical points for the Onsager functional. Phil. Trans. R. Soc. A 2021;379:20200110. http://dx.doi.org/10.1098/rsta.2020.0110
13. Manabe A, Bremer M, Karska M. Ferroelectric nematic phase at and below room temperature. Liq Cryst. 2021; 48: 1079-1086. https://doi.org/10.1080/02678292.2021.1921867.
14. Saha R, Nepal P, Feng C, Hossein MS, Gleeson JT, Sprunt S, Twieg RJ, Jakli A. Multiple ferroelectric nematic phases of a highly polar liquid crystal compound. 2021. arXiv:2104.06520.
15. Folcia CL, Ortega J, Vidal R, Sierra T, Etxebarria J. The ferroelectric nematic phase: An optimum liquid crystal candidate for nonlinear optics. 2021. arXiv: 2112.06040.
16. Humphries RL, James PG, Luckhurst GR. J Chem Soc Faraday Trans II. Molecular Field Treatment of Nematic Liquid Crystals. 1972; 68: 1031–1039. DOI: https://doi.org/10.1039/F29726801031.
17. Ypma JGY, Vertogen G, Koster HT. A Molecular Statistical Calculation of Pretransitional Effects in Nematic Liquid Crystals. Mol Cryst Liq Cryst. 1976; 37: 57-69. DOI: 10.1080/15421407608084346.
18. Madhusudana NV. Simple molecular model for ferroelectric nematic liquid crystals exhibited by small rodlike mesogens. Phys Rev E. 2021;104:014704. DOI: 10.1103/PhysRevE.104.014704.